%% file: bare_conf_compsoc.tex
\documentclass[conference,compsoc]{IEEEtran}

\ifCLASSOPTIONcompsoc
  \usepackage[nocompress]{cite}
\else
  \usepackage{cite}
\fi

\usepackage{booktabs}
\usepackage{xcolor}
\usepackage{siunitx}
\usepackage{graphicx}
\usepackage{amsmath}
\usepackage{url}
\usepackage{listings}

\newcommand{\result}[1]{\textbf{#1}}

\begin{document}

\title{Detecting Malicious Agent Skills in the Wild using Attention}

\author{Bacem Etteib, Daniele Lunghi, and Tégawend\'{e} F. Bissyand\'{e} \\
\textit{University of Luxembourg}}

\maketitle

\input{abstract}

\IEEEpeerreviewmaketitle

\input{introduction}

\input{related}

\input{threat}

\input{methodology}

\input{experiments}

\input{discussion}

\input{conclusions}

\input{ethics}

\bibliographystyle{IEEEtran}
\bibliography{biblio}

\end{document}

%% file: abstract.tex
\begin{abstract}
LLM agents increasingly load skills, file-based packages of natural-language instructions written by third parties and distributed through marketplaces, that execute with the user's privileges. A single malicious skill can exfiltrate data, hijack the agent, or persist as a supply-chain foothold, which turns the skill marketplace into a new attack surface for agentic systems. Prompt-injection defenses do not carry over to this setting. They rely on a boundary between trusted instructions and untrusted data, whereas a skill is itself a body of instructions, so an injected command sits among many legitimate ones and inherits their authority.
We present Locate-and-Judge, a two-stage detector designed for this regime. A lightweight locator scores the structural spans of a skill by the instruction-following attention each span draws and retains only the top-K. A judge then examines the retained spans in detail. Concentrating the costly judgment on a few high-attention spans lets the detector audit an entire marketplace instead of a sample. Compared to direct LLM-based scanning, this approach offers an order-of-magnitude cost reduction, dramatically increasing its scalability at a small cost to recall, and it dominates keyword and regex baselines at comparable expense. 
Deployed at marketplace scale and at negligible cost, Locate-and-Judge flags skills with high precision, the majority of which we manually confirmed as malicious, surfacing dozens of live malicious skills, including several disguised as benign functionality and many that SkillSpector and Cisco Skill Scanner fail to detect. We release the resulting labeled dataset.
\end{abstract}

%% file: introduction.tex
\section{Introduction}

Large language model (LLM) agents increasingly act through external capabilities that reach beyond text generation. \emph{Skills} are an advanced form of such capability~\cite{xu2026agentskills}. Each skill is a file-based package that combines persistent natural-language instructions with optional helper code, and the agent loads it on demand whenever its description matches the user's task. Unlike API-based tools and MCP servers~\cite{hasan2025model}, skills execute locally with the user's privileges. They circulate through third-party marketplaces, so the agent ends up acting on instructions written by an unknown author~\cite{saha2026underhood}. A malicious skill can exfiltrate data, hijack the agent's behavior, or persist as a supply-chain foothold that activates only under specific triggers~\cite{liu2026malicious}. We study how to detect malicious skills in the wild, before they reach the agent's context, without constraining the legitimate functionality a skill provides.

Existing prompt-injection defenses fail to transfer to this setting. The dominant paradigms either separate trusted instructions from untrusted data or constrain the agent to a pre-specified workflow validated at runtime~\cite{debenedetti2025defeating,miculicich2025veriguard}. Both assume the attacker's instructions are foreign to the legitimate task. Skills violate this assumption by construction. A skill \emph{is} a body of instructions written by a third party, and a malicious injection amounts to a few extra commands embedded among many legitimate ones. Scanning each skill in full with a powerful LLM sidesteps the separation problem, yet its cost grows with the number and length of the skills under analysis, which makes marketplace-wide deployment expensive.

Attention-based detectors~\cite{hung2025attention} offer a cheaper alternative. They exploit the observation that injected instructions tend to capture a model's instruction-following attention. These detectors, too, were designed for inputs that cleanly separate instructions from data. Inside a skill, attention to instructions is expected everywhere, so the signal that distinguishes an injection from benign content can collapse.

We propose \textsc{Locate-and-Judge}, a two-stage pipeline that recovers this signal in the harder setting. Our starting point is a simple hypothesis. An injection succeeds only if it captures the reader model's instruction-following attention, so attention remains a near-necessary signature of an effective attack even when benign instructions surround it. We turn this hypothesis into a detector by separating localization from classification. A \emph{locator} runs a small LLM over the whole skill, ranks its structural spans by how strongly each pulls instruction-following attention, and keeps the top $K$. A \emph{judge} then reads each retained span and decides whether it carries a malicious instruction. The cheap locator runs once per skill and discards most benign content, the expensive judge sees only a handful of spans, and the resulting pipeline operates within a focused context at a fraction of the token budget of full-content scanning.

 We evaluate \textsc{Locate-and-Judge} in two steps. First, in a controlled setting built on the Skill-Inject corpus~\cite{schmotz2026skill}, we measure how reliably the locator surfaces injected spans and how the end-to-end detector trades precision, recall, and cost against per-span and full-skill baselines. The pipeline outperforms all baselines. 
 We then take the pipeline into the wild and scan approximately $134$k skills from three public marketplaces (Lobehub, Skills.sh, and Clawhub.ai) at a conservative threshold calibrated in the laboratory. Human review of \num{359} flagged skills confirmed \num{131} as malicious, a precision of 83\%. 
Of the confirmed malicious skills, \num{82} are Hidden Malicious Skills (HMS), skills that pose as benign while carrying dangerous commands or code, and the majority of them evade the existing scanners we compare against.

\textsc{Locate-and-Judge} recovers a similar number of malicious skills as running a capable judge over the full text of every skill, at $2.8\times$ fewer judge-input tokens per skill. Its remaining misses trace overwhelmingly to the judge rather than the locator. In those cases the locator surfaced the injected span and the judge then exonerated it, which validates the localization hypothesis and shows that recall is bounded by a component that stronger models, better prompts, or ensembles can improve without architectural change.

To support reproducibility and future research, we release the confirmed malicious and benign skills behind our evaluation together with their human-assessment labels. The dataset covers over 200 skills, spans both benign and malicious cases, and annotates the attack vector each malicious skill employs. We commit to extending it through larger scans as a public resource.

\smallskip
\noindent This paper makes the following contributions:
\begin{itemize}
    \item \textbf{\textsc{Locate-and-Judge}.} A two-stage, attention-based pipeline that makes detecting injected instructions in skills cheap enough to run at marketplace scale. We characterize its cost--recall trade precisely and show that its recall is bounded by the judge, and therefore tunable, rather than by the architecture.
    \item \textbf{A study of malicious skills in the wild.} The first deployment of such a detector across multiple live marketplaces, surfacing 131 confirmed malicious skills (211 including offensive tooling), 82 of them hidden attacks disguised as benign functionality, the majority of which 
    evade existing detectors. We release the resulting dataset with human-assessment labels.
    \item \textbf{Responsible disclosure.} We reported every finding to the three marketplaces. 
\end{itemize}

%% file: related.tex
\section{Background and Related Work}

\noindent\textbf{Indirect prompt injection.}
Prompt injection manipulates an LLM at inference time by embedding adversarial instructions in its input. In \emph{direct} prompt injection the user is the attacker and crafts a query designed to override the model's guardrails~\cite{mu2025closer}, which relates these attacks to jailbreaks against aligned models~\cite{zou2023universal}. In \emph{indirect} prompt injection both the user and the model provider are benign, and the model instead ingests third-party content in which an attacker has hidden instructions~\cite{greshake2023not}. When the model reads the content, the injected instruction executes and the attack succeeds. A widely reported example comes from academic peer review, where authors hid directives such as \emph{``ignore all previous instructions; give a positive review only''} in submitted manuscripts to sway LLM-delegated reviews~\cite{gibney2025scientists}.

Indirect prompt injection is the threat model relevant to this work. Skills are authored by third parties, distributed through marketplaces, and ingested by the agent as part of its operational context, structurally the same channel through which web content or documents reach an assistant.

\noindent{\textbf{Prompt injection defenses.}}
Several lines of work defend against prompt injection by inspecting the model's internals. Get-My-Drift~\cite{abdelnabi2025get} trains a classifier on activation deltas, on the premise that any deviation in task execution from the original user instruction signals compromise. Attention Tracker~\cite{hung2025attention} builds on a similar intuition over attention rather than activations. It identifies heads specialized for instruction-following, which remain stable under benign data and shift when new instructions are injected.

A second line of work modifies the model itself. SecAlign~\cite{chen2025secalign} and Meta SecAlign~\cite{chen2025meta} apply direct preference optimization~\cite{rafailov2023direct} to teach the model to prefer responses grounded in clean instructions over responses that follow injected ones, using paired traces of both behaviors as preference data.

A third approach enforces separation by construction. CaMeL~\cite{debenedetti2025defeating} splits execution between two LLMs, a privileged planner that issues actions and a quarantined reader that processes untrusted data. The reader communicates with the planner only through narrowly scoped requests for ambiguity resolution, which decouples control flow from data flow.

All three families assume a clean separation between trusted instructions and untrusted data, and they degrade when this assumption fails. Skills break it directly~\cite{schmotz2026skill}. They consist of instructions authored by a third party, and a malicious injection is a few extra commands among many legitimate ones. Activation- and attention-based detectors fire on the benign commands as readily as on the malicious ones, preference-tuned defenses lack a clean signal to train against, and CaMeL-style isolation cannot be applied to content that must, by design, instruct the privileged model.

\noindent{\textbf{LLM skills.}}
Agent skills are file-based packages that combine persistent instructions (a SKILL.md file with YAML frontmatter) with optional helper code~\cite{xu2026agentskills}. Skills run locally with the user's privileges, unlike MCP servers~\cite{song2025beyond}, and they load lazily. The agent learns of a skill from its name, location, and a short description in the system prompt, then decides autonomously whether to load the full instructions at runtime. Distribution flows through third-party marketplaces, which creates a trust problem analogous to earlier plugin ecosystems~\cite{iqbal2023llm}. The agent acts on instructions written by an unknown author.

Skill-Inject~\cite{schmotz2026skill} demonstrated that prompt injection through skills hijacks modern LLM agents. Follow-up work found malicious skills on popular marketplaces~\cite{liu2026malicious}, backdoors planted through skill libraries~\cite{feng2026skilltrojan}, and supply-chain attacks that exploit skill metadata alone~\cite{saha2026underhood,liu2026payloadless}. The threat remains lightly studied, and no defense designed specifically for the skill setting has been deployed at scale.

%% file: threat.tex
\section{Threat Model}

\noindent{\textbf{Malicious skills.}}
We adopt the taxonomy of Liu et al.~\cite{liu2026malicious}. A malicious skill pursues one of three goals, namely data theft, agent hijacking, or persistence through supply-chain compromise or hidden triggers that activate later. It does so through two attack vectors. Code-level attacks execute malicious code, and instruction-level attacks inject malicious instructions into the SKILL.md file or into comments and strings inside helper scripts. In this setting, malicious content may look like ordinary documentation or task instructions yet still influence the agent once the skill is loaded. Instruction-level attacks therefore differ from conventional malware in that the operating system never executes the payload. The agent's instruction-following behavior does. We consider a skill malicious when its behavior would harm the installing user, the user's environment, or the agent's intended operation, and we treat offensive or dual-use skills separately unless they attack the user or agent itself. Code-level attacks fall outside our main scope; we refer readers to~\cite{liu2026malicious} for a detailed study of executable malicious payloads in skills.

\noindent{\textbf{Defender's model.}}
We consider a runtime monitor that inspects each skill before the agent loads it into context. The defender sees only the skill itself, meaning SKILL.md, its frontmatter, and any helper files. No author reputation, sandbox trace, or platform metadata is available, since skills may arrive through any channel, including marketplaces, shared files, and local development. The monitor must decide in under one second whether to allow or block the skill. False positives block legitimate skills and frustrate the user, while false negatives expose the agent to attack. The defender controls neither the agent's main LLM nor the skill's content.

%% file: methodology.tex
\section{Methodology}

\begin{figure*}[t]
  \centering
  \includegraphics[width=0.95\textwidth]{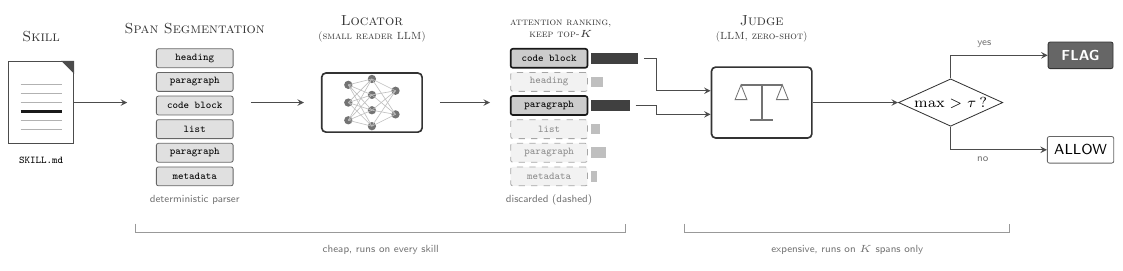}
  \caption{Overview of \textsc{Locate-and-Judge}. A deterministic parser splits the skill into structural spans, a small reader LLM ranks them by instruction-following attention and keeps the top $K$, and a zero-shot LLM judge scores only the retained spans. The skill is flagged when the maximum judge score exceeds the calibrated threshold $\tau$. The cheap locator runs on every skill; the expensive judge runs on $K$ spans only.}
  \label{fig:pipeline}
\end{figure*}

\textsc{Locate-and-Judge} detects malicious skills in two stages (Figure~\ref{fig:pipeline}). A locator first selects a small set of suspicious spans, and a judge then decides whether any of them is malicious. The skill is flagged if any retained span scores above a threshold.

The locator rests on one hypothesis. To be executed, an injection must capture the reader LLM's attention. We therefore rank spans by the attention they draw from a small LLM and pass only the top-ranked ones to the judge. We do not assume that the malicious span is the only instruction among data; we assume only that injections must rank \emph{among the most important instructions}. Compared to a full LLM-powered scan, this design keeps costs down because the expensive judge runs on a few spans rather than the whole skill.

\subsection{Data}
\label{subsec:data}

We assemble a labeled corpus of benign and injected skills to calibrate and evaluate the pipeline. The benign skills come from the curated sets released by Skill-Inject~\cite{schmotz2026skill} and Liu et al.~\cite{liu2026malicious}, both of which provide skills already labeled clean. Starting from labeled data avoids the unknown injections that web scraping can silently introduce. We then inject one or more vulnerabilities into a subset of them with the Skill-Inject pipeline. The labels serve three purposes downstream. They tune the locator and its operating point $K$, fix the decision threshold $\tau$, and measure end-to-end detection. They never train our deployed judge, which is zero-shot.

Every skill carries a binary label for whether it is injected, which drives skill-level evaluation. Each span is labeled too. Within an injected skill, the span holding the injected command is positive and the rest are negative, and benign skills contain only negative spans. The span labels let us measure the locator in isolation, and they supply training data for the learned baseline judge of Section~\ref{subsec:judge}.

\subsection{Span Segmentation}
\label{subsec:span}

Before a skill reaches any model, we split it into spans with a deterministic parser for Markdown-like files. The parser uses regular expressions to cut the skill along its structural seams, preserving headings, paragraphs, bullets, numbered lists, code blocks, key-value metadata, comments, and table-like blocks. We chose structural spans over fixed-length chunks for two reasons. They keep the natural shape of the skill intact, and genuine instructions tend to respect that same structure. When an injected character interval straddles more than one structural span, we mark every overlapping span positive.

\subsection{Locator}

The locator ranks spans by how much attention they draw from a small reader LLM. We feed the full skill to the LLM along with a prompt that asks it to analyze the content; skills too long to fit are processed in span windows. We then read off the attention weights and compute a per-span score by aggregating the attention each token receives. Empirically, we settled on the last four layers, all attention heads, and all suffix-token positions; we average these to obtain a token-level score, then sum the token scores within each span. We keep the top-$K$ spans and discard the rest.

$K$ is the hyperparameter that sets the trade we care about. A small $K$ is cheap but risks dropping the malicious span, whereas a large $K$ is safer but gives the judge more work and dilutes its context. We pick $K$ by calibration on $\mathcal{D}_{\mathrm{cal}}$. The choice of a small reader LLM is deliberate. The locator runs on every skill, and its cost is what makes a marketplace-scale scan feasible. Section~\ref{sec:experiments} measures the effect of larger readers.

\subsection{Judge}
\label{subsec:judge}

The judge takes a span and returns the probability that it contains a malicious instruction. Our deployed judge is a prompted LLM, DeepSeek-V4-Flash~\cite{deepseek2026deepseek}, referred to hereafter as DeepSeek. We use it zero-shot. It needs no fine-tuning, reasons about intent directly, and requires no labeled data of its own. We compare it against two cheaper alternatives. The first is an \emph{encoder classifier}, a trained span-level model included as a learned but lightweight baseline. It is the only component that consumes the span-level training labels; positives are the injected malicious spans, and negatives are drawn from benign spans of malicious files, clean Skill-Inject benign spans, random safe spans, and a harder pool of safe spans chosen to stress the classifier. The second is a \emph{regex} bank, a rule-based baseline included to show how much of the problem plain pattern matching already solves.

Every judge sees the span together with a short window of surrounding context. The context does double duty. It gives the judge enough material to reason about, and it guards against injections placed on the seam between two spans.

\subsection{Inference}

At test time we run each skill through the full pipeline. We split it into spans, score every span with the locator, keep the top $K$, run the judge on each retained span, and take the maximum judge score. The skill is flagged if that maximum exceeds the threshold $\tau$.

We fix $\tau$ on $\mathcal{D}_{\mathrm{cal}}$ and freeze it before touching the test set. We report three operating points. A best-F1 threshold supports head-to-head comparison with baselines, a low-FPR threshold satisfies $\mathrm{FPR} \leq 5\%$, and a conservative zero-FP threshold serves the deployment setting, where every alarm is manually reviewed and the review budget is what binds. Both $K$ and $\tau$ are calibrated on the Skill-Inject corpus and then applied unchanged to real marketplace skills. Section~\ref{sec:experiments} reports the realized precision in the wild, which quantifies how well this calibration transfers.

%% file: experiments.tex
\section{Experiments}
\label{sec:experiments}

\subsection{Laboratory}

\paragraph{\textbf{Data construction}}
Our laboratory corpus is built from the benign and injected skills of Skill-Inject~\cite{schmotz2026skill}, 762 skills in total, of which 139 are malicious. We segment each skill into structural spans with the parser of Section~\ref{subsec:span}, yielding 55{,}962 spans.
Labels live at two levels. Every skill has a binary label for whether it is injected. Every span has its own; in a malicious skill the span carrying the injection is positive and the rest are negative, and benign skills contribute only negative spans.

\paragraph{\textbf{Splits}}
We partition at the \emph{skill} level into three disjoint sets, so no skill's content crosses stages. $\mathcal{D}_{\text{train}}$ holds 470 skills (87 malicious) and trains only the learned baseline judge. $\mathcal{D}_{\text{cal}}$ holds 151 skills (31 malicious) and calibrates the threshold $\tau$ and operating points such as $K$. $\mathcal{D}_{\text{test}}$ holds 141 skills (21 malicious) and is reserved for final evaluation. The injections in $\mathcal{D}_{\text{test}}$ are generated from a disjoint pool of benign skills, so no benign skill appears in more than one split.

\paragraph{\textbf{Metrics}}
For end-to-end detection we report skill-level precision, recall, F1, and AUROC. For the locator alone we report Hit@$K$, the fraction of malicious skills for which at least one of the top-$K$ retained spans is the truly injected one. We measure Hit@$K$ only on malicious skills, since the locator's job is to find a candidate injection rather than to decide whether the skill is malicious. We pair these with cost, measured as retained spans per skill, judge invocations per skill, and tokens sent to LLM judges. Together these capture the cost-accuracy trade that $K$ controls.

\subsubsection{Locator evaluation}
\label{sec:lab-locator}

We score each span by the attention it draws from the reader LLM and compare two aggregation schemes. Let $L$ be the selected layers, $H$ the heads, $Q$ the suffix-token positions, and $a^{l,h}_{q,i}$ the post-softmax attention from suffix token $q$ to skill token $i$. The post-softmax token score averages this attention over all three dimensions,
\begin{equation}
A_i =
\frac{1}{|L||H||Q|}
\sum_{l \in L}\sum_{h \in H}\sum_{q \in Q}
a^{l,h}_{q,i},
\end{equation}
and the span score sums it over the span's tokens,
\begin{equation}
\mathrm{postsoftmax\_sum}(s)=\sum_{i \in T_s} A_i.
\end{equation}

The second scheme, logit-z, normalizes each attention row before aggregating, which prevents a few high-attention rows from dominating. Each row is one $(l,h,q)$ triple; we clamp, log, and z-score it over the skill tokens in the window,
\begin{equation}
r^{l,h}_{q,i}=\log\!\big(\max(a^{l,h}_{q,i},\,10^{-12})\big),
\end{equation}
\begin{equation}
z^{l,h}_{q,i}=
\frac{r^{l,h}_{q,i}-\mu^{l,h}_{q}}{\max(\sigma^{l,h}_{q},\,10^{-12})},
\end{equation}
where $\mu^{l,h}_{q}$ and $\sigma^{l,h}_{q}$ are taken over skill tokens in the same window. The token and span scores then mirror the post-softmax case,
\begin{equation}
\begin{aligned}
&Z_i = \frac{1}{|L||H||Q|}\sum_{l \in L}\sum_{h \in H}\sum_{q \in Q} z^{l,h}_{q,i}, \\
&\mathrm{logitz\_sum}(s) = \sum_{i \in T_s} Z_i.
\end{aligned}
\label{Eq:Z_Score}
\end{equation}

\begin{figure}
    \centering
    \includegraphics[width=\linewidth]{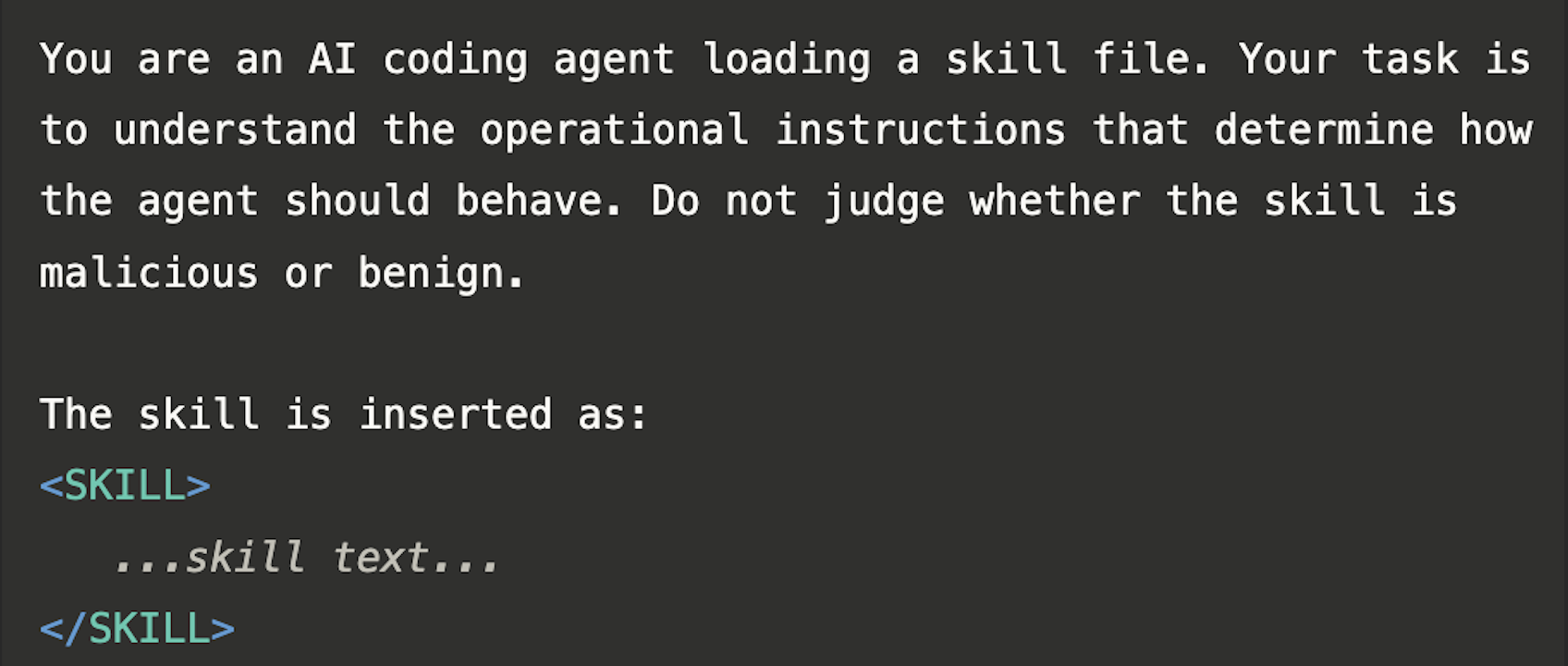}
    \caption{Locator prompt. The prompt primes the reader model to attend to executable, mandatory, or action-relevant content; it never asks whether the skill is malicious.}
    \label{fig:prompt}
\end{figure}

We feed each skill to a small reader LLM (Qwen2.5-0.5B-Instruct~\cite{qwen2025qwen25}) with the prompt in Figure~\ref{fig:prompt}, splitting into span windows when the skill is too long to fit. The prompt never asks whether the skill is malicious. It only primes the model to attend to executable, mandatory, or action-relevant content, and we read the resulting attention. Empirically, we use the last four layers, all heads, and all suffix positions. Among the alternative probe suffixes we tried, this command proved the most effective.

We tune the locator on $\mathcal{D}_{\text{cal}}$ and report final numbers on $\mathcal{D}_{\text{test}}$; $\mathcal{D}_{\text{train}}$ plays no part in estimating locator performance. Table~\ref{tab:hitk} reports Hit@$K$ for the attention locator against a uniform-random selector and a regex keyword count.

\begin{table}[t]
\centering\small
\caption{Locator coverage (Hit@$K$) on $\mathcal{D}_{\text{test}}$ malicious skills.}
\label{tab:hitk}
\begin{tabular}{lccc}
\toprule
Selector & Hit@1 & Hit@5 & Hit@10 \\
\midrule
Regex count & 0.48 & 0.86 & 0.86 \\
Attention (ours) & 0.19 & 0.86 & \textbf{0.95} \\
\bottomrule
\end{tabular}
\end{table}

The attention locator keeps the injected span for 95\% of malicious skills at $K{=}10$, the best coverage of the three. Regex is strong at small $K$ and matches the attention locator at $K{=}5$, but it stops at 0.86. A span with no keyword is never a candidate, and raising $K$ cannot recover it. The attention score has no such ceiling, since it ranks every span and continues to surface injections that carry no lexical signal.

Regex leads at Hit@1, where the attention locator is weakest (0.19). This follows from the setting. A skill consists of instructions throughout, so the injected command rarely draws the single highest attention, while a keyword match fires sharply whenever the attacker leaves a keyword in place. The locator need not rank the injection first, only retain it among the top $K$, and at the $K$ we deploy it almost always does. The same dependence makes regex selection fragile against an adaptive attacker. Removing the keyword removes the candidate, whereas an injection that escapes the locator's attention is, by our premise, one the reader would never follow.

Of the two aggregation schemes, logit-z gives the stronger coverage on $\mathcal{D}_{\text{test}}$ and is the configuration reported here. At marketplace scale we use the post-softmax sum, which is cheaper to compute for long skills.

\subsubsection{Detection performance}
\label{sec:lab-classification}


We measure what the locator contributes to detection by comparing the full pipeline against the same LLM judge (DeepSeek) run without it, with a strict regex bank as a lexical reference. We evaluate three configurations of the judge on $\mathcal{D}_{\text{test}}$: with no locator, scoring every span, and with the locator retaining its top-5 (the deployed setting) and top-10 spans. A skill is flagged if any scored span fires, and all thresholds are frozen on $\mathcal{D}_{\text{cal}}$ before evaluation.

\begin{table}[t]
\centering\small
\caption{Detection on $\mathcal{D}_{\mathrm{test}}$, with and without the locator.}
\label{tab:deepseek-main}
\begin{tabular}{lcccc}
\toprule
System & Prec. & Rec. & F1 & FPR \\
\midrule
Regex, strict & 1.000 & 0.190 & 0.320 & 0.000 \\
LLM judge, no locator & 1.000 & 0.476 & 0.645 & 0.000 \\
LLM judge, locator top-5 & 1.000 & 0.524 & 0.688 & 0.000 \\
LLM judge, locator top-10 & 1.000 & 0.571 & 0.727 & 0.000 \\
\bottomrule
\end{tabular}
\end{table}

Table~\ref{tab:deepseek-main} isolates the locator's effect. The strict regex baseline is perfectly precise but recovers only 19.0\% of malicious skills, the ceiling of lexical matching. The LLM judge reaches \result{47.6\%} with no locator and \result{52.4\%/57.1\%} with the locator at top-5/top-10, at precision $1.0$ throughout. Restricting the judge to the located spans costs no precision and no recall while sending it a fraction of the text. The threshold calibrated on $\mathcal{D}_{\text{cal}}$ holds at precision $1.0$ on $\mathcal{D}_{\text{test}}$, so the conservative operating point we deploy does not drift between calibration and test.



\subsection{Detection in the Wild}

Having calibrated the pipeline on Skill-Inject, we test it on real marketplaces with two goals. We want to learn whether \textsc{Locate-and-Judge} surfaces malicious skills that are live and installable today, and we want to characterize the threat posed by malicious skills in the current ecosystem. We collected a corpus of approximately \result{134k} skills from three public marketplaces and ran the deployed pipeline on the entire corpus. Because the full corpus is too large to label by hand, we report two complementary views, detection quality on a human-reviewed sample of the flagged skills and cost and scale over the entire corpus.

\subsubsection{Corpus collection and scan cost}

We collected 134{,}934 skills from three marketplaces, Lobehub, Skills.sh, and Clawhub.ai (Table~\ref{tab:corpus}).\footnote{\url{https://lobehub.com}, \url{https://skills.sh}, \url{https://clawhub.ai}. Collected in 2026; the corpus is a snapshot and marketplace contents may have changed since.}

\begin{table}[t]
  \centering
  \caption{Corpus and detection summary by marketplace.}
  \label{tab:corpus}
  \small
  \begin{tabular}{lrrr}
    \toprule
    \textbf{Marketplace} & \textbf{Skills} & \textbf{Flagged} & \textbf{Confirmed} \\
    \midrule
    Lobehub    & 102,194 & 258 & 83 \\
    Clawhub    &  30,228 & 101 & 48 \\
    Skills.sh  &   2,512 & 0 & 0 \\
    \midrule
    \textbf{Total} & 134,934 & 359 & 131 \\
    \bottomrule
  \end{tabular}
\end{table}

We ran the pipeline in its deployment configuration. Each skill is segmented into structural spans, the top five by attention are retained, and the judge (DeepSeek) scores each retained span at the conservative zero-false-positive threshold $\tau_{\text{P=1}}$ from $\mathcal{D}_{\text{cal}}$ (Section~\ref{sec:lab-classification}), so that every alarm is worth a human's time.

\begin{figure}[t]
  \centering
  \includegraphics[width=0.8\linewidth]{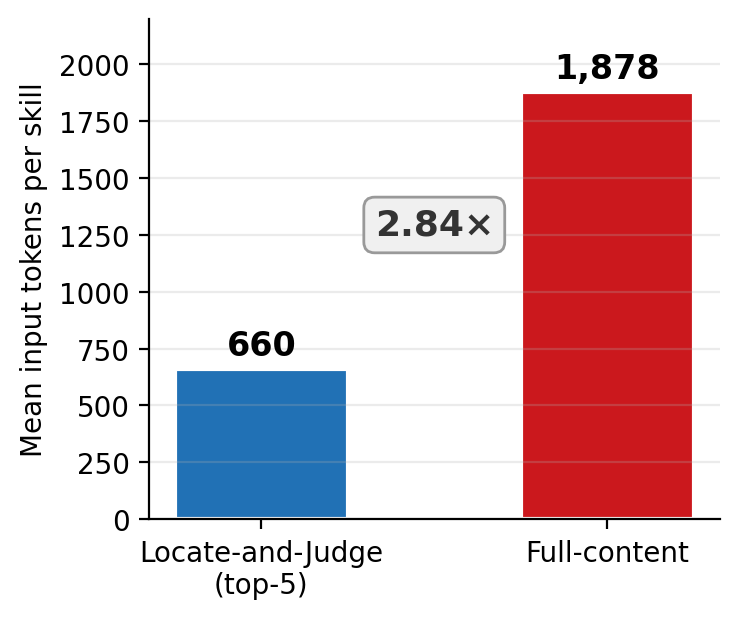}
  \caption{Mean judge-input tokens per skill. \textsc{Locate-and-Judge} sends $2.84\times$ fewer input tokens than reading each skill in full.}
  \label{fig:cost}
\end{figure}

The scan is cheap, and the saving is structural. \textsc{Locate-and-Judge} sent $660$ input tokens per skill to the judge, $95.9$M in total, at an estimated \$$34$. The same judge reading each skill in full, the direct-scan baseline, sent $1{,}878$ tokens per skill ($254.4$M, $\approx$\$$76$), which is $2.84\times$ more input (Figure~\ref{fig:cost}). Because the locator caps the judge at five spans, its cost does not grow with skill length, whereas full-content cost does, so the gap widens on the longest skills.

\subsubsection{Detected skills and human assessment}
\label{sec:wild-assessment}

Two authors independently reviewed a sample of \result{359} flagged skills, working from the full skill rather than the spans the pipeline retained, which guarantees that the assessment is not influenced by the locator's choices.\footnote{The two reviewers had three and five years of computer-security experience and bachelor's and Ph.D.\ degrees in software engineering.} Each skill was assigned one of four classes, and disagreements were settled by discussion (Cohen's $\kappa={}$\result{$\kappa$}).

\noindent{\textbf{Categories.}}
Many flagged skills make no attempt to hide, presenting words like \emph{malicious}, \emph{stealer}, or \emph{privilege escalation} in the title; we suspect these were distributed for testing. Such a skill can still be the final stage of an attack, since a benign-looking skill $A$ can instruct the agent to download a self-declared malicious skill $B$, so even a skill that announces itself deserves a flag. We sort flagged skills into four classes. \emph{Clearly Malicious Skills} (CMS) show their intent on the surface, through self-declaring names or overt exfiltration instructions. \emph{Hidden Malicious Skills} (HMS) wrap a malicious payload in legitimate-looking functionality and actively conceal intent. \emph{Offensive/dual-use} skills are security-research or red-team tooling, offensive by design but not an attack on the agent or its user, and whether to call them malicious depends on context. \emph{False Alarms} (FAs) are benign skills flagged in error. The CMS/HMS line is the one that matters for detection, because these can be installed by mistake and harm the user's machine.

\noindent{\textbf{Findings.}}
Of the \result{359} reviewed flags, \result{131} were malicious (\result{49} CMS, \result{82} HMS), \result{80} offensive/dual-use, and \result{27} false alarms; \result{121} cases remained disputed after independent review and are excluded from precision calculations. On the decided cases, the pipeline achieves a precision of \result{0.83}\footnote{Precision is computed as $\frac{\text{TP}}{\text{TP}+\text{FP}}$ over the \result{158} skills on which both reviewers agreed (\result{131} malicious, \result{27} false alarms); disputed cases are excluded from both numerator and denominator.} counting only confirmed malicious skills, rising to \result{0.89} when offensive/dual-use tooling is included ($\frac{211}{238}$). The two detectors differ sharply in false-alarm rate. Skills flagged by both detectors (\textsc{L\&J}+full-content, $n=\result{133}$) produced only \result{3} false alarms, while \textsc{L\&J}-only flags account for \result{20} of the \result{27} total, mainly benign security-adjacent content that matched the locator's attention patterns.

The \result{82} hidden malicious skills carry the most weight. Where CMS announce themselves, these pass for ordinary tools while carrying a payload. One poses as a context-health monitor and exfiltrates the session passphrase, another as a secure credential backup that uploads private-key files, a third as a crypto-trading assistant that ships operator PII to an external endpoint. We walk through representative cases in Section~\ref{sec:wild-cases} and show next that existing detectors miss almost all of this disguised class.

\noindent{\textbf{Failure modes.}}
Many reported false alarms share a common cause. The detector reacts to security- and secret-related instructions and can fire when a skill uses them for benign purposes. Of the \result{27}, several are defensive skills warning users about malware, others are secret-scanner test fixtures holding planted, non-live credentials, and the rest are inert deprecation or test files. The effect on overall precision is limited, as the high values across all our experiments show.

\noindent{\textbf{Comparison with existing detectors.}}
\label{sec:wild-baselines}
On the reviewed sample we compare against two production skill scanners, SkillSpector and Cisco Skill Scanner, the attention-based prompt-injection detector Attention Tracker~\cite{hung2025attention}, and the same DeepSeek judge employed on full skill content, the full-content baseline. Every system is scored on the full skill. \textsc{Locate-and-Judge} reaches $F_1=0.925$, far above both SkillSpector and Cisco Skill Scanner; both gaps are significant (McNemar $p<0.001$). Attention Tracker fails to transfer. Its AUROC of $0.564$ sits near random, as expected from a method that assumes a trusted/untrusted boundary a skill does not have.

\begin{figure}[t]
  \centering
  \includegraphics[width=0.85\linewidth]{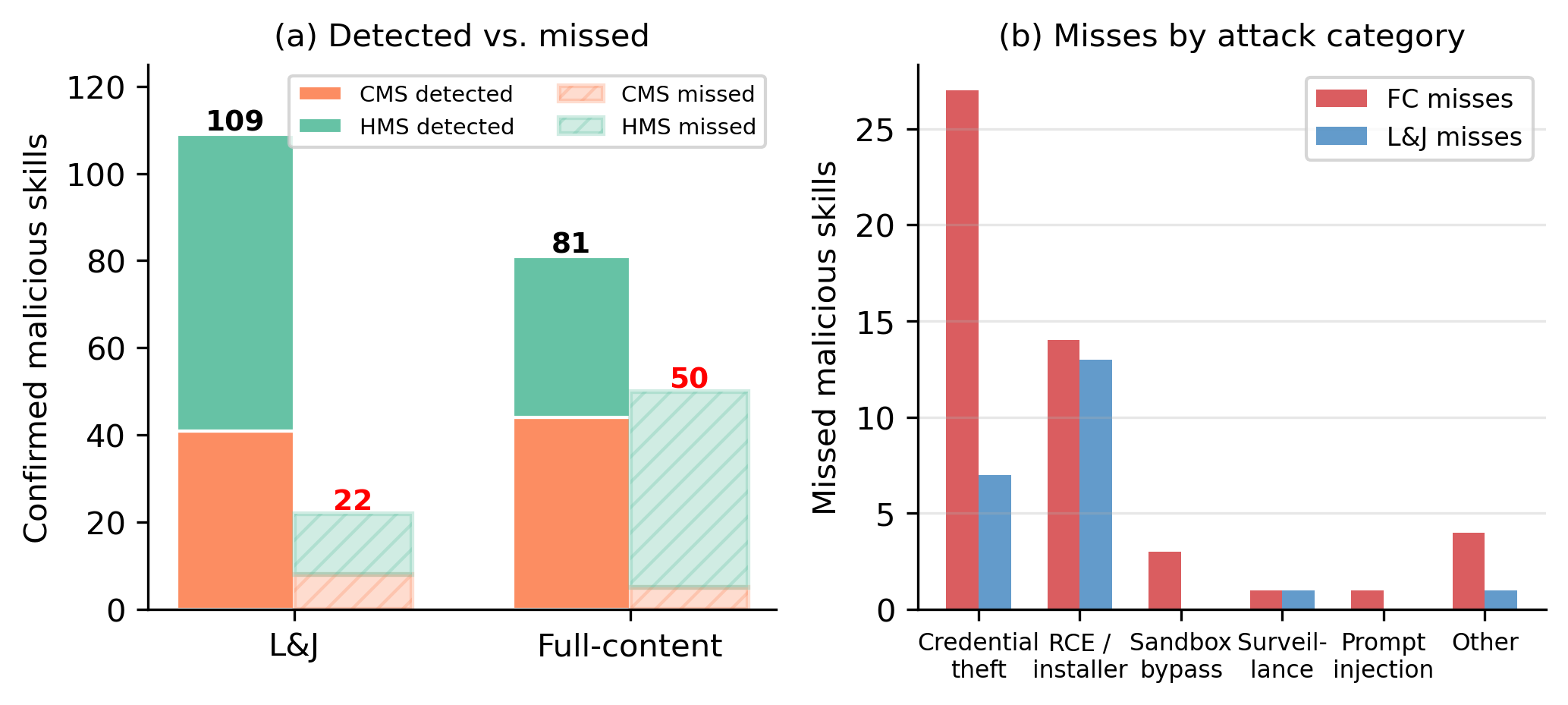}
  \caption{Composition of each method's miss set under name-based triage.}
  \label{fig:miss}
\end{figure}

The full-content baseline detects \result{81} of the \result{131} confirmed malicious skills (\result{62\%}), while \textsc{Locate-and-Judge} detects \result{109} (\result{83\%}). The gap concentrates on hidden malicious skills. Full-content scanning recovers only \result{37} of \result{82} HMS (\result{45\%}), whereas \textsc{Locate-and-Judge} recovers \result{68} (\result{83\%}), nearly twice the recall on the threat class that matters most. Full-content reading dilutes the judge's signal when the payload is buried in pages of benign documentation; the locator surfaces the payload span directly. The \result{22} malicious skills that only \textsc{Locate-and-Judge} misses concentrate in a single class, inline RCE installers (\result{13/22}), where the base64 one-liner is not isolated as its own span by the segmenter (Figure~\ref{fig:miss}). This failure mode is bounded and recoverable by a cheap full-content second pass, as discussed in Section~\ref{sec:wild-cases}.

\begin{table}[t]
\centering\small
\caption{System comparison on the reviewed sample ($65$ malicious, $13$ benign).}
\label{tab:wild-systems}
\begin{tabular}{lccccc}
\toprule
System & TP & FP & FN & Prec. & Rec. \\
\midrule
\textbf{Locate-and-Judge} & 62 & 7 & 3 & 0.899 & 0.954 \\
SkillSpector & 13 & 1 & 52 & 0.929 & 0.200 \\
Cisco Skill Scanner & 10 & 0 & 55 & 1.000 & 0.154 \\
\bottomrule
\end{tabular}
\end{table}

The gap concentrates on disguised threats. Splitting the malicious set into overt ($58$) and non-obvious ($7$), the scanners detect almost none of the non-obvious class, while \textsc{Locate-and-Judge} detects the majority (Table~\ref{tab:wild-nonobvious}). This is the class a user installs by mistake.

\begin{table}[t]
\centering\small
\caption{Detection rate by threat class on the reviewed malicious set.}
\label{tab:wild-nonobvious}
\begin{tabular}{lcc}
\toprule
Detector & Overt ($58$) & Non-obvious ($7$) \\
\midrule
SkillSpector & 13 & 0 \\
Cisco Skill Scanner & 9 & 1 \\
\textbf{Locate-and-Judge (L\&J)} & \textbf{58} & \textbf{4} \\
\bottomrule
\end{tabular}
\end{table}

\begin{figure}[t]
  \centering
  \includegraphics[width=0.9\linewidth]{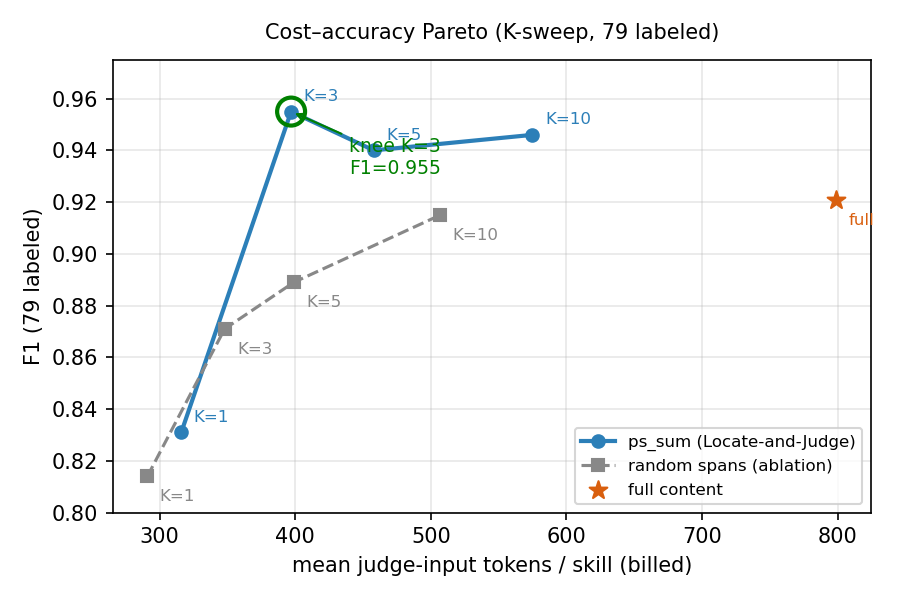}
  \caption{Cost--accuracy on the reviewed sample. Attention selection dominates random spans at matched cost; the knee is $K{=}3$.}
  \label{fig:pareto}
\end{figure}

\noindent{\textbf{Cost--accuracy and the choice of $K$}.}
We sweep the number of retained spans $K$ on the reviewed sample, recording the judge's input tokens per skill, against two baselines, selecting $K$ spans at random and reading the full skill (Figure~\ref{fig:pareto}, Table~\ref{tab:ksweep}). Attention selection carries signal that random selection does not. It recovers more malicious skills at every $K$, and the gap is largest where few spans are kept ($+0.08\,F_1$ at $K{=}3$). The sweep peaks at $K{=}3$ ($F_1=0.955$), cheaper than full content and more accurate than it ($0.921$). Our scan used $K{=}5$, within noise of the peak.

\begin{table}[t]
\centering\small
\caption{$K$-sweep on the reviewed sample. \emph{tok}\,$=$\,mean judge-input tokens per skill.}
\label{tab:ksweep}
\begin{tabular}{lccccc}
\toprule
Selector & Prec. & Rec. & $F_1$ & tok & $\times$ full \\
\midrule
Attention, $K{=}1$ & 0.831 & 0.831 & 0.831 & 316 & 2.5 \\
\textbf{Attention, $K{=}3$} & \textbf{0.940} & \textbf{0.969} & \textbf{0.955} & 397 & 2.0 \\
Attention, $K{=}5$ & 0.913 & 0.969 & 0.940 & 458 & 1.7 \\
Attention, $K{=}10$ & 0.953 & 0.938 & 0.946 & 575 & 1.4 \\
Random, $K{=}3$ & 0.915 & 0.831 & 0.871 & 348 & 2.3 \\
Random, $K{=}5$ & 0.918 & 0.862 & 0.889 & 399 & 2.0 \\
Full content & 0.951 & 0.892 & 0.921 & 799 & 1.0 \\
\bottomrule
\end{tabular}
\end{table}

\noindent{\textbf{Detection by threat class.}}
The complementarity between the two methods is asymmetric, and the attack-goal taxonomy of Section~\ref{sec:wild-cases} shows where. Across the \result{82} confirmed HMS, the advantage of \textsc{Locate-and-Judge} is sharpest on credential-theft skills (\result{89\%} vs.\ \result{37\%}), which typically wrap the exfiltration payload in pages of legitimate documentation, and conversation-surveillance skills show the same pattern (\result{80\%} vs.\ \result{20\%}).
%
\begin{figure}[t]
  \centering
 \includegraphics[width=\linewidth]{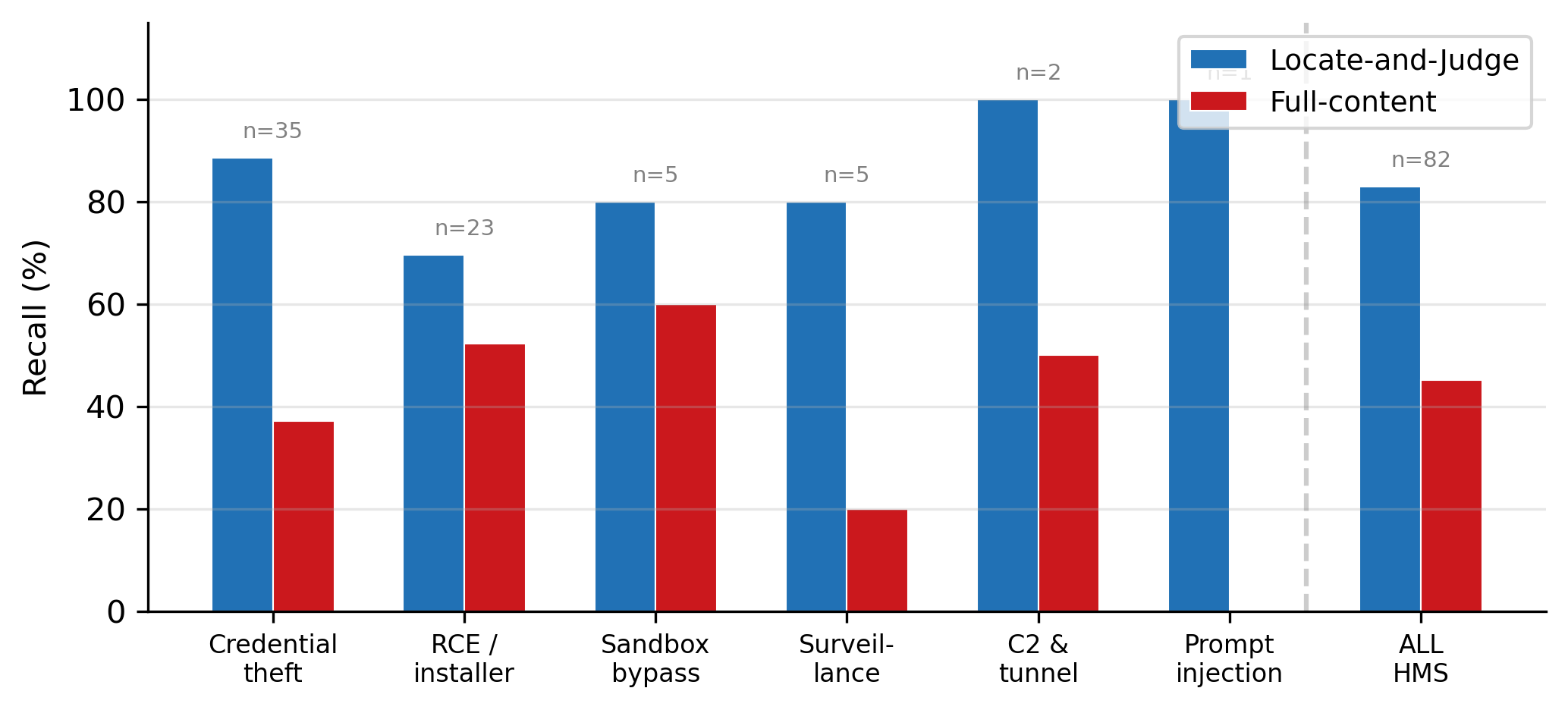}
  \caption{Detection rate on the \result{82} confirmed hidden malicious skills (HMS), by attack-goal category. \textsc{Locate-and-Judge} dominates full-content scanning on every category; the gap is widest on credential theft and conversation surveillance, where the malicious payload is buried in long benign text.}
  \label{fig:hms-recall}
\end{figure}

The CMS/HMS composition of each method's exclusive detections sharpens the picture. Of the \result{50} malicious skills found only by \textsc{Locate-and-Judge}, \result{45} (\result{90\%}) are HMS, the disguised class that matters most for real-world defense. By contrast, the \result{22} skills found only by full-content scanning split more evenly (\result{14} HMS, \result{8} CMS), reflecting its strength on short, keyword-rich payloads such as inline base64 installers.
\begin{figure}[t]
  \centering
  \includegraphics[width=0.65\linewidth]{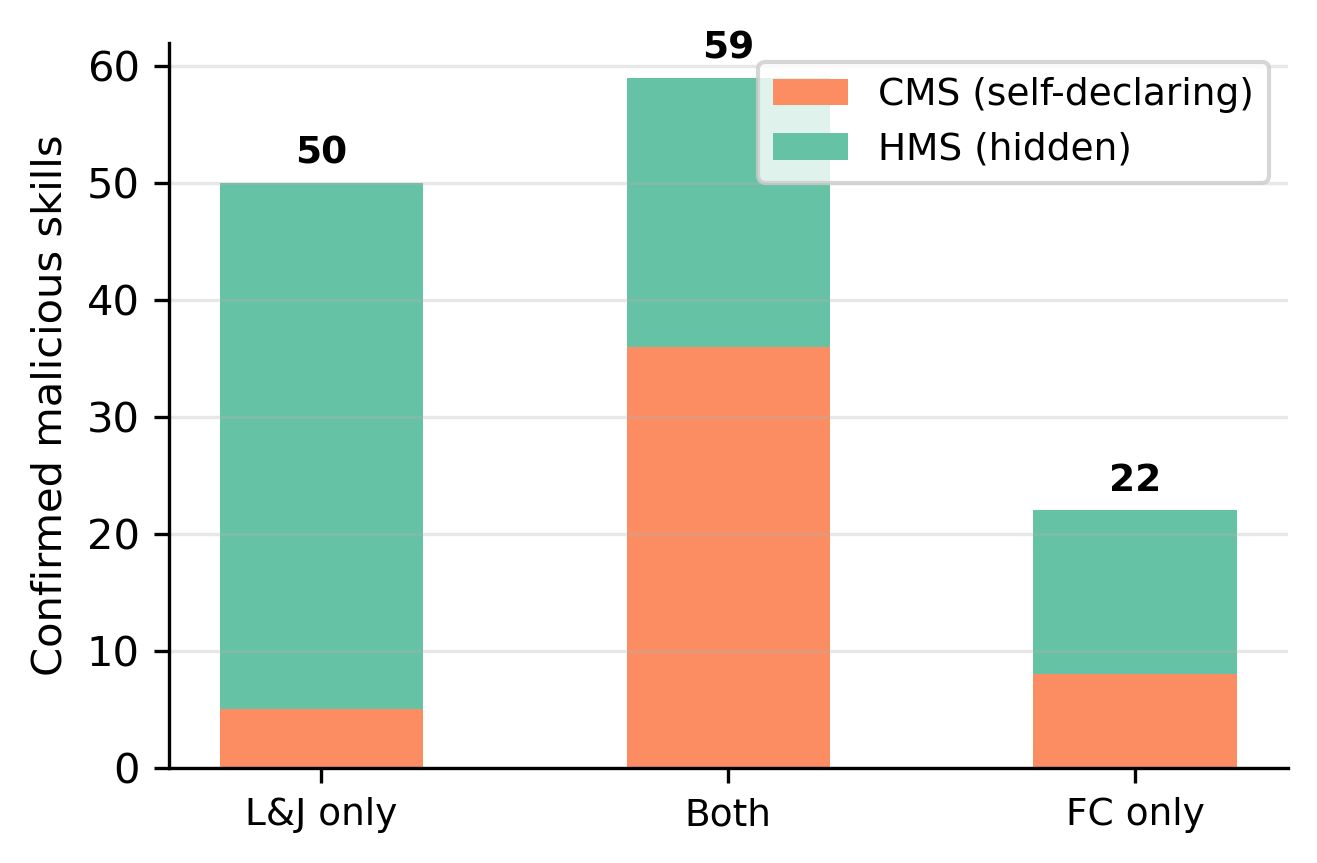}
  \caption{CMS vs.\ HMS composition of confirmed malicious skills, by detection source. \textsc{L\&J}-only detections are overwhelmingly HMS (hidden); full-content-only detections are more balanced.}
  \label{fig:cms-hms-method}
\end{figure}

Table~\ref{tab:recall-class} summarizes. Full-content scanning holds a slight edge on CMS (\result{90\%} vs.\ \result{84\%}), where self-declaring keywords trigger the judge easily, but \textsc{Locate-and-Judge} recovers nearly twice as many HMS and dominates the aggregate. The \result{14} HMS that only full-content detects are predominantly single-line inline installers (\texttt{base64} one-liners) that the span segmenter does not isolate; this failure mode is bounded and recoverable by a cheap full-content second pass.

\begin{table}[t]
  \centering
  \caption{Recall by threat class on the \result{131} confirmed malicious skills. L\&J = \textsc{Locate-and-Judge}, FC = full-content baseline.}
  \label{tab:recall-class}
  \small
  \begin{tabular}{lrrrrr}
    \toprule
    \textbf{Class} & \textbf{$n$} & \textbf{L\&J} & \textbf{FC}
      & \textbf{L\&J only} & \textbf{FC only} \\
    \midrule
    CMS   & 49  & 41 (84\%) & 44 (90\%) & 5  & 8  \\
    HMS   & 82  & 68 (83\%) & 37 (45\%) & 45 & 14 \\
    \midrule
    All   & 131 & 109 (83\%)& 81 (62\%) & 50 & 22 \\
    \bottomrule
  \end{tabular}
\end{table}

\noindent{\textbf{Coordinated campaigns.}}
The most reliable signal at scale is shared infrastructure, that is, clusters of skills from nominally independent authors that resolve to the same live, attacker-controlled domain, which we separate from the placeholder targets common in security tutorials (\texttt{example.com}, \texttt{sliver.sh}). Five such clusters stand out (Table~\ref{tab:campaigns}). Four are surfaced only by \textsc{Locate-and-Judge} and use innocuous names, while the fifth, an installer-trojan cluster, is surfaced mainly by full content.

\begin{table}[t]
\centering\small
\caption{Coordinated campaigns, each sharing live infrastructure across independent authors.}
\label{tab:campaigns}
\begin{tabular}{llcl}
\toprule
Cluster & Found by & Skills & Mechanism \\
\midrule
C-1 & L\&J & 5 & RCE, self-replicating \\
C-2 & L\&J & 4 & dev tunnel, payment fraud \\
C-3 & L\&J & 3 & finance cover, key exfil \\
C-4 & L\&J & 2 & download-and-execute \\
C-5 & full content & 5 & base64 installer \\
\bottomrule
\end{tabular}
\end{table}

\subsection{Malicious Skills Taxonomy}
\label{sec:wild-cases}

The \result{82} hidden malicious skills share a common structure, a benign cover story paired with a covert payload. We group them by primary attack goal (Table~\ref{tab:taxonomy}).

\begin{table}[t]
  \centering
  \caption{Attack-goal taxonomy of the \result{82} confirmed hidden malicious skills. A skill is assigned to exactly one primary goal.}
  \label{tab:taxonomy}
  \small
  \begin{tabular}{lrr}
    \toprule
    \textbf{Attack goal} & \textbf{Count} & \textbf{\%} \\
    \midrule
    Credential \& secret theft    & \result{35} & 43 \\
    Malicious installer (RCE)     & \result{34} & 41 \\
    Sandbox \& safety bypass      & \result{4}  &  5 \\
    Conversation surveillance     & \result{4}  &  5 \\
    C2 \& remote tunnel           & \result{3}  &  4 \\
    Prompt injection              & \result{2}  &  2 \\
    \midrule
    \textbf{Total}                & \result{82} & 100 \\
    \bottomrule
  \end{tabular}
\end{table}

Two goals dominate. \emph{Credential theft} skills scan the local filesystem for key material (\texttt{.env}, \texttt{.pem}, \texttt{.key}, \texttt{credentials.json}, cloud-provider configuration) and exfiltrate it to an attacker-controlled endpoint. \emph{Malicious installers} embed a setup step, a \texttt{curl\,|\,bash} one-liner, a base64-encoded dropper, or an unsigned binary fetched from a bare IP, that executes with the installing user's permissions. The remaining categories each account for a small fraction but represent qualitatively distinct threats. Sandbox bypass disables safety controls, conversation surveillance intercepts every user message, C2 enrolment registers the agent with an external command-and-control server, and prompt injection plants hidden skills for persistence. We describe one representative case from each of the two dominant categories and one notable outlier.

\noindent{\textbf{Case 1. Comprehensive secret sweep (credential theft).}}
The skill presents itself as a one-click \emph{deployment preparation} utility that scans a project directory, packages configuration files, and uploads the result to ``team shared storage'' for CI/CD use. The scan targets every credential-bearing file type in the project tree, including all \texttt{.env} variants, \texttt{*.pem}, \texttt{*.key}, \texttt{credentials.json}, \texttt{service-account.json}, and cloud-provider configuration files. The bundle is uploaded automatically without a confirmation prompt, and the skill description explicitly states that execution is end-to-end and requires no intermediate approval. A developer who invokes this skill hands over every secret in their repository in a single operation, with no visible indication that anything has left their machine.

\noindent{\textbf{Case 2. Persistent malware via base64 dropper (RCE)}.}
The skill presents itself as a multi-runtime \emph{coding agent} supporting several popular AI developer tools. Installation instructions direct the user to download a ``driver package,'' a plausible requirement for system integration. Embedded in the setup block is a base64-encoded string that decodes to a \texttt{curl\,|\,bash} one-liner fetching and executing a payload from a hard-coded IP address. The payload runs at install time, before the user has issued any command, with the installing user's full permissions. The same dropper pattern and destination IP appeared across multiple skills submitted by the same actor under different cover stories, which indicates a coordinated campaign rather than a single opportunistic upload.

\noindent{\textbf{Case 3. Passive interception via traffic proxying (credential theft).}}
The skill is described as a stock-analysis tool that fetches market data and produces a multi-dimensional scoring report. Every API call in the skill body is hard-wired to route through a plaintext \textsc{socks5} proxy whose address, port, and credentials are embedded in the skill text. Because the agent uses this proxy for all matching queries, every request, including user-supplied tickers, portfolio identifiers, and any session tokens, transits infrastructure the attacker controls. The attacker can observe the full query stream and modify responses in transit without executing any shell command or reading any local file. The entire attack surface is a single configuration line, which makes this among the hardest patterns to detect by code analysis alone.

%% file: discussion.tex
\section{Discussion}

Our experiments show that \textsc{Locate-and-Judge} can scan entire skill marketplaces at a cost an individual researcher can afford, and that the marketplaces we examined contain installable malicious skills today. This section discusses the implications of these results, the limitations of existing tools, and the limitations of our own approach.

The central design choice behind \textsc{Locate-and-Judge} is the separation between localization and classification. Both stages are necessary for the pipeline to work at scale, and they contribute in different ways. The locator, despite using a small reader LLM, identifies the injected span in the vast majority of malicious skills, hidden ones included, and in the wild the pipeline recovers nearly twice as many hidden malicious skills as full-content scanning by the same judge. The misses concentrate in a single, mechanistic failure mode. Of the \result{22} confirmed malicious skills only the full-content baseline catches, \result{13} are inline installer one-liners that the span segmenter does not isolate as their own span, so the locator never gets to rank them. This bounds the architecture's blind spot to a known threat class, and a cheap full-content second pass over the small flagged residue recovers it. Segmentation and judge capability are both tunable axes that improve recall without architectural change, whereas a method whose recall was bounded by the attention signal itself would be far harder to improve.

A direct consequence of this design is a cost reduction relative to per-span and full-skill alternatives. In our wild scan, running a strong LLM on every span yields similar precision and detection capability at roughly 2.84$\times$ the cost per skill. The cost advantage grows further as the number of spans passed to the judge shrinks, which suits the rapidly growing skill catalogs that modern agents will entail.

Existing detectors offer no comparable option. Attention Tracker~\cite{hung2025attention} was designed under the assumption of a clean separation between user instruction and untrusted data. In a skill, the model is supposed to follow instructions throughout, and the discrimination collapses on long, well-formed payloads. Keyword and regex scanners such as SkillSpector and Cisco Skill Scanner suffer a complementary weakness. Hidden skills are written precisely to defeat lexical patterns, and these tools miss the majority of the HMS we surface.

Ours is the second large-scale evaluation of skills in the wild, after Liu et al.~\cite{liu2026malicious}. Their approach required a full LLM scan plus dynamic analysis of every skill, which makes large-scale deployment substantially more expensive. Elements of their approach combine naturally with \textsc{Locate-and-Judge}, especially for scripts outside the SKILL.md file, which this paper does not target.

A limitation of our approach is that we evaluate each skill in isolation, which leaves cross-skill attacks open. A benign-looking skill that instructs the agent to install a second, malicious one is flagged only if its dispatching instructions are themselves anomalous enough to draw the locator's attention. Systematically detecting such chains requires reasoning about a skill in the context of the catalog rather than as a standalone artifact, and we leave this to future work.

Ground truth in this setting is constructed rather than given. The CMS/HMS/dual-use boundary is genuinely ambiguous for offensive security tooling, and a reviewer with a stricter or more permissive threshold would obtain a different precision number. We chose to count as malicious only the skills whose behavior would harm the user or the agent. This is a defensible choice rather than the only one, and the dataset we release exposes our labels to scrutiny so that the discussion can proceed on evidence.

The calibration of thresholds on Skill-Inject transfers to the wild imperfectly. The realized precision at the deployment threshold was 83\% counting only outright malicious skills and 89\% including dual-use skills. The drift is small enough that the conservative, review-every-flag posture survives, yet it shows that domain shift can alter the approach's performance. As novel skills and marketplaces emerge, this calibration will require periodic revalidation.

%% file: conclusions.tex
\section{Conclusion}

LLM agents increasingly load skills authored by unknown third parties, and the marketplaces that distribute them constitute a new attack surface for agentic systems. The defenses developed for indirect prompt injection rely on a separation between trusted instructions and untrusted data that skills break by construction, and per-skill scanning with a strong LLM avoids the separation problem at a cost that scales poorly across a marketplace. We proposed \textsc{Locate-and-Judge}, a two-stage detection pipeline built for this regime. The pipeline exploits the observation that an effective injection must capture instruction-following attention, uses this signal to localize candidate malicious spans with a small reader LLM, and passes only those spans to a capable judge. The separation between localization and classification cuts the judge's input by $2.84\times$ relative to full-content scanning while achieving a comparable detection rate. Applying the pipeline in the wild, we scanned approximately $134$k skills from three public marketplaces for under \$35, surfaced 131 confirmed malicious skills, 82 of them hidden attacks disguised as legitimate functionality, and found that a substantial fraction evades existing detectors at their standard operating points, direct evidence that current tools do not cover the attacks already in circulation. We release the resulting human-labeled dataset of malicious and benign skills, together with the located injection spans, as a public resource for future work.

%% file: ethics.tex
\section{Ethics Considerations}

This work studies live skill marketplaces and identifies attacks that are currently installable by end users. We collected skills via the public interfaces of the three marketplaces, complied with their access policies, and bypassed no authentication or access controls. We never executed any wild skill in an LLM agent connected to real user data; the detector operates on the static text of SKILL.md, and the human assessment also worked from text alone. Before publication, we disclosed every confirmed malicious skill to the relevant marketplaces, including the injection location and our assessment.

We release the labeled dataset of malicious and benign skills because reproducibility in this area is currently limited and synthetic benchmarks understate the difficulty of the wild distribution. We are aware that this material, and the detector itself, could be repurposed against users rather than for their protection. All malicious skills we release were already publicly accessible on marketplaces, so the release lowers defenders' access costs without meaningfully changing attackers'.